\documentclass[11pt,floatfix,showpacs]{revtex4}
\usepackage[dvips]{graphicx,color}
\usepackage{amsmath}
\usepackage{amsfonts}
\usepackage[latin1]{inputenc}

\begin{document}

\title{Mass dependence of the deconfinement and chiral
restoration critical temperatures in nonlocal SU(2) PNJL models}

\author{V. Pagura$^{a,b}$, D. G\'omez Dumm$^{b,c}$ and N.N.\ Scoccola$^{a,b,d}$}

\address{$^{a}$ Physics Department, Comisi\'on Nacional de
Energ\'{\i}a At\'omica, Av.Libertador 8250, 1429 Buenos Aires,
Argentina \\
$^{b}$ CONICET, Rivadavia 1917, 1033 Buenos Aires, Argentina \\
$^{c}$ IFLP, CONICET $-$ Dpto.\ de F\'{\i}sica, Universidad
Nacional de La Plata, C.C. 67, 1900 La Plata, Argentina,\\
$^{d}$ Universidad Favaloro, Sol{\'\i}s 453, 1078 Buenos Aires,
Argentina}

\begin{abstract}
In the framework of nonlocal SU(2) chiral quark models with
Polyakov loop, we analyze the dependence of the deconfinement and
chiral restoration critical temperatures on the explicit chiral
symmetry breaking driven by the current quark mass. Our results
are compared with those obtained within the standard local
Polyakov$-$Nambu$-$Jona-Lasinio (PNJL) model and with lattice QCD
calculations. For a wide range of pion masses, it is found that
both deconfinement and chiral restoration critical temperatures
turn out to be strongly entangled, in contrast with the
corresponding results within the PNJL model. In addition, it is
seen that the growth of the critical temperatures with the pion
mass above the physical point is basically linear, with a slope
parameter which is close to the existing lattice QCD estimates. On
the other hand, within the present mean field calculation we find
an early onset of the first order transition expected in the large
quark mass limit.
\end{abstract}

\pacs{12.39.Ki, 11.30.Rd, 12.38.Mh}

\maketitle

\section{Introduction}

It is widely believed that as the temperature and/or density
increase, strongly interacting matter undergoes some kind of
transition from a hadronic phase, in which chiral symmetry is
broken and quarks are confined, to a partonic phase in which
chiral symmetry is restored and/or quarks are deconfined. The
detailed understanding of this phenomenon is relevant not only in
particle physics but also e.g.\ in the the study of the early
universe, the interior of neutron stars, etc., therefore it has
become an issue of great interest in recent years, both
theoretically and experimentally~\cite{Rischke:2003mt}. From the
theoretical point of view, one way to address this problem is
through lattice QCD calculations~\cite{All03,Fod04,Kar03}.
However, even if significant improvements have been done in this
field in the last years, this ab initio approach is not yet able
to provide a full understanding of the QCD phase diagram. One
serious difficulty in this sense is given by the so-called sign
problem, which prevents straightforward simulations at finite
baryon density. In this situation it is worth to develop
alternative approaches, such as the study of effective models that
show consistency with lattice QCD results and can be extrapolated
into regions not accessible by lattice techniques. Here we will
concentrate on one particular class of effective theories, namely
the so-called nonlocal Polyakov$-$Nambu$-$Jona-Lasinio (nlPNJL)
models~\cite{Blaschke:2007np,Contrera:2007wu,Hell:2008cc,Contrera:2012wj}, in
which quarks move in a background color field and interact through
covariant nonlocal chirally symmetric four point couplings.
Related Polyakov-Dyson-Schwinger equation models have also been
recently analyzed~\cite{Horvatic:2010md}. These approaches, which
can be considered as an improvement over the (local) PNJL
model~\cite{Meisinger:1995ih,Fukushima:2003fw,Megias:2004hj,Ratti:2005jh,
Roessner:2006,Mukherjee:2006hq,Sasaki:2006ww}, offer a common
framework to study both the chiral restoration and deconfinement
transitions. In fact, the nonlocal character of the interactions
arises naturally in the context of several successful approaches
to low-energy quark dynamics~\cite{Schafer:1996wv,RW94}, and leads
to a momentum dependence in the quark propagator that can be made
consistent~\cite{Noguera:2008} with lattice
results~\cite{Parappilly:2005ei,Furui:2006ks}. Moreover, it has
been found that, under certain conditions, it is possible to
derive the main features of nlPNJL models starting directly from
QCD~\cite{Kondo:2010ts}. From the phenomenological side, it has
been shown~\cite{BB95,BGR02,Scarpettini:2003fj,GomezDumm:2006vz}
that nonlocal models provide a satisfactory description of hadron
properties at zero temperature and density.

As mentioned above, it is important to consider situations in which the
results obtained within effective models can be compared with available
lattice QCD calculations. For example, it is clear that vacuum properties
such as the pion mass and decay constant, as well as other features related
to the chiral/deconfinement transitions (like e.g.\ the nature of the
transitions, or the critical temperatures) will depend on basic parameters
of QCD, such as the number of quark flavors and the values of current quark
masses $m_q$. In particular, for the simplified case of two degenerate
flavors with $m_{u}=m_{d}=m$, the dependence of several relevant quantities
on $m$ has been studied with some detail in lattice QCD. Thus, the
corresponding analysis within nlPNJL models can provide an interesting test
of the reliability of this effective approach. Actually, it has already been
shown that several chiral effective
models~\cite{Berges:1997eu,Dumitru:2003cf,Braun:2005fj} are not able to
reproduce the behavior of the critical temperatures observed in lattice QCD
when one varies the parameters that explicitly break chiral symmetry (i.e.\
the current quark masses, or the pion mass in the case of meson models) at
vanishing chemical potential. This fact has been taken as an indication that
the transition may be not just dominated by pure chiral
dynamics~\cite{Fraga:2008be}. It is worth to notice that in the framework of
the standard (local) NJL model the enhancement of the critical temperature
with $m$ is too strong in comparison with lattice QCD estimates. Although
the inclusion of confinement effects through the coupling to the Polyakov
loop weakens this enhancement, one finds a too large splitting between the
chiral restoration and deconfinement transition
temperatures~\cite{Kahara:2009sq}. The presence of confinement effects
together with a strong entanglement between the chiral restoration and
deconfinement transitions is indeed one of the features of nlPNJL
models~\cite{Pagura:2011rt}.

In view of the above mentioned points, the aim of the present work is to
study the effect of explicit chiral symmetry breaking on the deconfinement
and chiral restoration critical temperatures within nlPNJL models. This
article is organized as follows. In Sec.\ II we provide a description of the
model, proposing two alternative parameterizations. In Sec.\ III we analyze
the $m$-dependence of some pion properties and compare the results with
existing lattice calculations. In Sec.\ IV we analyze the current quark mass
dependence of the critical temperatures at vanishing chemical potential,
comparing our results with those obtained in alternative models and lattice
QCD. Finally in Sec.\ V we summarize our main results and conclusions.

\section{Formalism}

We consider a nonlocal SU(2) chiral quark model that includes quark
couplings to the color gauge fields. The corresponding Euclidean effective
action is given by~\cite{Contrera:2010kz}
\begin{equation}
S_{E}= \int d^{4}x\ \left\{
\bar{\psi}(x)\left( -i\gamma_{\mu}D_{\mu}
+\hat{m}\right)  \psi(x) -\frac{G_{S}}{2} \Big[ j_{a}(x)j_{a}(x)- j_{P}%
(x)j_{P}(x)\Big]+ \ {\cal U}\,(\Phi[A(x)])\right\}  \ , \label{action}%
\end{equation}
where $\psi$ is the $N_{f}=2$ fermion doublet $\psi\equiv(u,d)^T$, and
$\hat{m}={\rm diag}(m_{u},m_{d})$ is the current quark mass matrix. In
what follows we consider isospin symmetry, $m_{u}=m_{d}=m$. The fermion
kinetic term in Eq.~(\ref{action}) includes a covariant derivative
$D_\mu\equiv\partial_\mu - iA_\mu$, where $A_\mu$ are color gauge fields,
and the operator $\gamma_\mu\partial_\mu$ in Euclidean space is defined as
$\vec \gamma \cdot \vec \nabla + \gamma_4\partial/\partial \tau$,
with $\gamma_4=i\gamma_0$. The nonlocal currents $j_{a}(x),j_{P}(x)$ are
given by
\begin{align}
j_{a}(x)  &  =\int d^{4}z\ {\cal G}(z)\ \bar{\psi}\left(  x+\frac{z}{2}\right)
\ \Gamma_{a}\ \psi\left(  x-\frac{z}{2}\right)  \ ,\nonumber\\
j_{P}(x)  &  =\int d^{4}z\ {\cal F}(z)\ \bar{\psi}\left(  x+\frac{z}{2}\right)
\ \frac{i {\overleftrightarrow{\rlap/\partial}}}{2\ \kappa_{p}}
\ \psi\left(  x-\frac{z}{2}\right)\ ,
\label{currents}%
\end{align}
where,
$\Gamma_{a}=(\leavevmode\hbox{\small1\kern-3.8pt\normalsize1},i\gamma
_{5}\vec{\tau})$ and $u(x^{\prime}){\overleftrightarrow{\partial}%
}v(x)=u(x^{\prime})\partial_{x}v(x)-\partial_{x^{\prime}}u(x^{\prime})v(x)$.
The functions ${\cal G}(z)$ and ${\cal F}(z)$ in Eq.~(\ref{currents}) are
nonlocal covariant form factors characterizing the corresponding
interactions. Notice that the four currents $j_a(x)$ require a common form
factor ${\cal G}(z)$ in order to guarantee chiral invariance, while the
coupling $j_{P}(x)j_{P}(x)$ is self-invariant under chiral transformations.
The scalar-isoscalar component of the $j_{a}(x)$ current will generate a
momentum dependent quark mass in the quark propagator, while the
``momentum'' current $j_{P}(x)$ will be responsible for a momentum dependent
quark wave function renormalization. Now we perform a bosonization of the
theory, introducing bosonic fields $\sigma_{1,2}(x)$ and $\pi_a(x)$, and
integrating out the quark fields. Details of this procedure as well as of
the determination of vacuum and meson properties at vanishing temperature in
this framework can be found e.g.\ in Ref.~\cite{Noguera:2008}.

Since we are interested in the deconfinement and chiral restoration critical
temperatures, we extend the bosonized effective action to finite temperature
$T$. This can be done by using the standard Matsubara formalism. Concerning
the gauge fields $A_\mu$, we assume that quarks move on a constant
background field $\phi = A_4 = i A_0 = i g\,\delta_{\mu 0}\, G^\mu_a
\lambda^a/2$, where $G^\mu_a$ are SU(3) color gauge fields. Then the traced
Polyakov loop, which in the infinite quark mass limit can be taken as an
order parameter of confinement, is given by $\Phi=\frac{1}{3} {\rm Tr}\,
\exp( i \phi/T)$. We work in the so-called Polyakov gauge, in which the
matrix $\phi$ is given a diagonal representation $\phi = \phi_3 \lambda_3 +
\phi_8 \lambda_8$. This leaves only two independent variables, $\phi_3$ and
$\phi_8$. In the case of vanishing chemical potential, owing to the charge
conjugation properties of the QCD lagrangian, the mean field traced Polyakov
loop is expected to be a real quantity. Since $\phi_3$ and $\phi_8$ have to
be real valued, this condition implies $\phi_8=0$. The mean field traced
Polyakov loop reads then $ \Phi = \Phi^* = \left[ 1 + 2 \,\cos
\left(\phi_3/T\right)\right]/3$. Thus in the mean field approximation, which
will be used throughout this work, the thermodynamical potential
$\Omega^{\rm MFA}$ at finite temperature and zero chemical potential is
given by
\begin{align}
\Omega^{\rm MFA} =  \,- \,4 T \sum_{c=r,g,b} \ \sum_{n=-\infty}^{\infty}
\int \frac{d^3\vec p}{(2\pi)^3} \ \log \left[ \frac{ (\rho_{n,
\vec{p}}^c)^2 + M^2(\rho_{n,\vec{p}}^c)}{Z^2(\rho_{n, \vec{p}}^c)}\right]+
\frac{\bar\sigma_1^2 + \kappa_p^2\; \bar\sigma_2^2}{2\,G_S} +
{\cal{U}}(\Phi ,\Phi^*,T) \ , \label{granp}
\end{align}
where $M(p)$ and $Z(p)$ are given by
\begin{eqnarray}
M(p) =  Z(p) \left[m_q + \bar\sigma_1 \ g(p) \right] \ , \qquad
Z(p) =  \left[ 1 - \bar\sigma_2 \ f(p) \right]^{-1}\ .
\label{mz}
\end{eqnarray}
Here $\bar\sigma_{1,2}$ are the mean field values of the scalar fields (note
that  $\bar \pi_a = 0$), while and $f(p)$, $g(p)$ are Fourier transforms of
${\cal F}(z)$ and ${\cal G}(z)$, respectively. We have also defined
\begin{equation}
\Big({\rho_{n,\vec{p}}^c} \Big)^2 =
\Big[ (2 n +1 )\pi  T + \phi_c \Big]^2 + {\vec{p}}\ \! ^2 \ ,
\end{equation}
where the quantities $\phi_c$ are given by the relation $\phi = {\rm
diag}(\phi_r,\phi_g,\phi_b) = {\rm diag}(\phi_3,-\phi_3,0)$.

To proceed we need to specify the explicit form of the Polyakov loop
effective potential ${\cal{U}}(\Phi ,\Phi^*,T)$. We consider two
alternative functional forms commonly used in the literature. The first
one, based on a Ginzburg-Landau ansatz, reads~\cite{Ratti:2005jh}
\begin{eqnarray}
{\cal{U}}_{\rm poly}(\Phi ,\Phi^*,T) =
T^4
\left[
-\,\frac{b_2(T)}{4}\,
\left( |\Phi|^2 + |\Phi^*|^2 \right)
-\,\frac{b_3}{6}\,
\left( \Phi^3 + (\Phi^*)^3 \right)
+\,\frac{b_4}{16}\,
\left( |\Phi|^2 + |\Phi^*|^2 \right)^2
\right]\ ,
\end{eqnarray}
where
\begin{eqnarray}
b_2(T) = a_0 +a_1 \left(\dfrac{T_0}{T}\right) + a_2\left(\dfrac{T_0}{T}\right)^2
+ a_3\left(\dfrac{T_0}{T}\right)^3\ .
\label{pol}
\end{eqnarray}
The potential parameters can be fitted to pure gauge lattice QCD data so as
to properly reproduce the corresponding equation of state and Polyakov loop
behavior. This yields~\cite{Ratti:2005jh}
\begin{eqnarray}
& & a_0 = 6.75\ ,\qquad a_1 = -1.95\ ,\qquad a_2 = 2.625\ , \nonumber \\
& & a_3 = -7.44 \ ,\qquad b_3 = 0.75 \ ,\qquad b_4 = 7.5 \ .
\end{eqnarray}
A second usual form is based on the logarithmic expression of the Haar
measure associated with the SU(3) color group integration. The potential
reads in this case~\cite{Roessner:2006}
\begin{equation}
{\cal{U}}_{\rm log}(\Phi ,\Phi^*,T) =
\left\{-\,\frac{1}{2}\, a(T)\,\Phi \Phi^* \;+
\;b(T)\, \log\left[1 - 6\, \Phi \Phi^* + 4\, \Phi^3 + 4\, (\Phi^*)^3
- 3\, (\Phi \Phi^*)^2\right]\right\}\; T^4 \ ,
\end{equation}
where the coefficients are parameterized as
\begin{equation}
a(T) = a_0 +a_1 \left(\dfrac{T_0}{T}\right) + a_2\left(\dfrac{T_0}{T}\right)^2
\ ,
\qquad
b(T) = b_3\left(\dfrac{T_0}{T}\right)^3 \ .
\label{log}
\end{equation}
Once again the values of the constants can be fitted to pure gauge lattice
QCD results. This leads to~\cite{Roessner:2006}
\begin{equation}
a_0 = 3.51\ ,\qquad a_1 = -2.47\ ,\qquad a_2 = 15.2\ ,\qquad b_3 = -1.75\ .
\end{equation}
The dimensionful parameter $T_0$ in Eqs.~(\ref{pol}) and (\ref{log})
corresponds in principle to the deconfinement transition temperature in
the pure Yang-Mills theory, $T_0 = 270$~MeV. However, it has been argued
that in the presence of light dynamical quarks this temperature scale
should be adequately reduced~\cite{Schaefer:2007pw}.

Finally, one has to take into account that $\Omega^{\rm MFA}$ turns out to
be divergent, thus it has to be regularized. Here we use the prescription
described e.g.\ in Ref.~\cite{GomezDumm:2004sr}, namely
\begin{equation}
\Omega^{\rm MFA}_{\rm reg} = \Omega^{\rm MFA} - \Omega^{\rm free} + \Omega^{\rm
free}_{\rm reg} + \Omega_0 \ , \label{omegareg}
\end{equation}
where $\Omega^{\rm free}$ is obtained from Eq.~(\ref{granp}) by setting
$\bar\sigma_1 = \bar\sigma_2=0$, and $\Omega^{\rm free}_{\rm reg}$ is the
regularized expression for the quark thermodynamical potential in the
absence of the four point fermion interaction,
\begin{equation}
\Omega^{\rm free}_{\rm reg} \ = \ -4 T \int \frac{d^3 \vec{p}}{(2\pi)^3}\;
\sum_{c=r,g,b} \ \sum_{s=\pm 1}\mbox{Re}\;
\ln\left[ 1 + \exp\left(-\;\frac{\epsilon_p + i s \phi_c}{T}
\right)\right]
\ ,
\label{freeomegareg}
\end{equation}
with $\epsilon_p = \sqrt{\vec{p}^{\;2}+m^2}\;$. The last term in
Eq.~(\ref{omegareg}) is just a constant fixed by the condition that
$\Omega^{\rm MFA}_{\rm reg}$ vanishes at $T=0$.

Given the full form of the thermodynamical potential, the mean field
values $\bar\sigma_{1,2}$ and $\phi_{3}$ can be obtained as solutions of
the coupled set of ``gap equations''
\begin{equation}
\frac{\partial \Omega^{\rm MFA}_{\rm reg}}
{\left(\partial\sigma_{1},\partial\sigma_{2}, \partial\phi_3\right)}\ = \ 0 \ .
\label{fullgeq}
\end{equation}
Once these mean field values are obtained, the behavior of other relevant
quantities as functions of the temperature and chemical potential can be
determined. We concentrate in particular in the chiral quark condensate
$\langle\bar{q}q\rangle = \partial\Omega^{\rm MFA}_{\rm reg}/\partial m$ and
the traced Polyakov loop $\Phi$, which will be taken as order parameters of
the chiral restoration and deconfinement transitions, respectively. The
associated susceptibilities will be defined as $\chi_{\rm ch}  =
\partial\,\langle\bar qq\rangle/\partial m$ and $\chi_{\rm PL} = d \Phi / d
T$.

In order to fully specify the model under consideration we proceed to fix
the model parameters as well as the nonlocal form factors $g(q)$ and $f(q)$
at the physical point $m_\pi = m_\pi^{\rm phys} = 139$~MeV. We consider two
different functional dependences for the form factors. The first one
corresponds to the often used exponential functions
\begin{equation}
g(q)= \mbox{exp}\left(-q^{2}/\Lambda_{0}^{2}\right) \ ,
\qquad f(q)= \mbox{exp}\left(-q^{2}/\Lambda_{1}^{2}\right)\ ,
\label{regulators}
\end{equation}
which guarantee a fast ultraviolet convergence of the loop integrals. Note
that the range (in momentum space) of the nonlocality in each channel is
determined by the parameters $\Lambda_0$ and $\Lambda_1$, respectively.
Fixing the current quark mass and chiral quark condensate at $T=\mu=0$ to
the phenomenologically adequate values $m = 5.7$ MeV and
$\langle\bar{q}q\rangle^{1/3} = 240$ MeV, the rest of the parameters can
be determined so as to reproduce the physical values of $f_\pi$ and
$m_\pi$, and by requiring $Z(0) = 0.7$, which is within the range of
values suggested by recent lattice
calculations~\cite{Parappilly:2005ei,Furui:2006ks}. In what follows this
choice of model parameters and form factors will be referred to as S1. The
second type of form factor functional forms considered here is given by
\begin{eqnarray}
g(q)  = \frac{1+\alpha_z}{1+\alpha_z\ f_z(q)} \frac{\alpha_m \ f_m (q) -m\
\alpha_z f_z(q)} {\alpha_m - m \ \alpha_z } \ , \qquad f(q)  = \frac{ 1+
\alpha_z}{1+\alpha_z \ f_z(q)} f_z(q)\ ,
\label{regulators_set2}
\end{eqnarray}
where
\begin{equation}
f_{m}(q) = \left[ 1+ \left( q^{2}/\Lambda_{0}^{2}\right)^{3/2} \right]^{-1}
\qquad , \qquad
f_{z}(q) = \left[ 1+ \left( q^{2}/\Lambda_{1}^{2}\right) \right]^{-5/2}.
\label{parametrization_set2}
\end{equation}
As shown in Ref.~\cite{Noguera:2008}, taking $m = 2.37$ MeV, $\alpha_m =
309$ MeV, $\alpha_{z}=-0.3$, $\Lambda_0 = 850$ MeV and $\Lambda_1 = 1400$
MeV one can very well reproduce the momentum dependence of mass and wave
function renormalization obtained in lattice calculations, as well as the
physical values of $m_\pi$ and $f_\pi$. In what follows this choice of
model parameters and form factors will be referred to as S2. Details on
the model parameters and the predictions for several meson properties in
vacuum can be found in Ref.~\cite{Noguera:2008}.

\section{Zero temperature pseudoscalar mass and decay constant away from the physical point}

As stated, we want to study the dependence of nlPNJL model
predictions on the amount of explicit chiral symmetry breaking.
This can be addressed by varying the current quark mass $m$, while
keeping the rest of the model parameters fixed at their values at
the physical point. As a first step we analyze in this section the
corresponding behavior of the pion mass and decay constant at
vanishing temperature, in comparison with that obtained in the
(local) NJL model and in lattice QCD. Our results are shown in
Fig.~1. As it is usual in lattice QCD literature, we choose to
take $m_\pi$ instead of $m$ as the independent variable in the
plots. The main reason for this is that $m_\pi$ is an observable,
i.e.\ a scale independent quantity, whereas $m$ is scale
dependent, hence its value is subject to possible ambiguities
related to the choice of the renormalization point. Dashed and
solid lines correspond to parameter sets S1 and S2, respectively,
while dotted lines correspond to the curves obtained within the
NJL model. Fat dots stand for lattice QCD results from
Ref.~\cite{Noaki:2008iy}. The upper panel shows the behavior of
the ratio $m_\pi^2/m$ as a function of $m_\pi$. In order to
account for the above mentioned renormalization point ambiguities,
the corresponding quark masses have been normalized so as to yield
the lattice value $m_{u,d}^{\rm \overline{MS}}\simeq 4.452$~MeV at
the physical point~\cite{Noaki:2008iy}. From the figure one
observes that both NJL and nlPNJL models reproduce qualitatively
the results from lattice QCD, showing a particularly good
agreement in the case of the nlPNJL model for parameter set S2.
However, the situation is different in the case of $f_\pi$ (lower
panel in Fig.~1): while the predictions from nonlocal models
follow a steady increase with $m_\pi$, in agreement with lattice
results, the local NJL model badly fails to reproduce this
behavior. Moreover, it can be seen that the discrepancy cannot be
cured even if one allows the coupling $G_S$ to depend on the
current quark mass (we have taken $G_S$ as a constant in nlPNJL
models)~\cite{Kahara:2009sq}. Thus, these results can be
considered as a further indication in favor of the inclusion of
nonlocal interactions as a step towards a more realistic
description of low momenta QCD dynamics.

\section{Dependence of critical temperatures on explicit chiral
symmetry breaking}

In this section we analyze within our nonlocal models the mass
dependence of the critical temperatures for the deconfinement and
chiral restoration transitions at vanishing chemical potential. We
start by considering the temperature dependence of the chiral and
deconfinement order parameters, as well as the corresponding
susceptibilities, for some representative values of the pion mass.
The corresponding results for the lattice motivated
parameterization S2 are shown in Fig.~\ref{fig2}, including both
the case of the polynomic (left panels) and logarithmic (right
panels) Polyakov potentials. Qualitatively similar results are
found for the exponential parameterization S1. Let us first
discuss the results for the polynomic potential. From the figure
it is seen that both transitions proceed as smooth crossovers, as
expected from lattice QCD results. Moreover, we observe that as
$m_\pi$ increases, the position of the peaks of the
susceptibilities $\chi_{\rm ch}$ and $\chi_{\rm PL}$ (left lower
panel) move simultaneously towards higher values of $T$, the
difference between the corresponding critical temperatures being
in all cases at the level of a few MeV. It is also seen that as
$m_\pi$ increases the chiral restoration transition tends to be
less pronounced, while the confinement one becomes steeper. In the
case of the logarithmic potential, we also observe that the
transition temperatures increase with $m_\pi$, as expected.
However, for a given value of $m_\pi$ both the chiral restoration
and deconfinement transitions are steeper than in the case of the
polynomic potential, and the correlation between them is stronger
(e.g.\ the difference between the transition temperatures for
$m_\pi = m_\pi^{\rm phys}$ is now about 0.02~MeV). In fact, it
turns out that already for $m_\pi=500$~MeV one finds a first order
phase transition. From lattice QCD results the onset of a first
order phase transition is indeed expected above a certain critical
amount of explicit symmetry breaking~\cite{Laermann:2003cv}.
However, present
estimations~\cite{Alexandrou:1998wv,Saito:2011fs,Fromm:2011qi}
indicate that the corresponding critical pseudoscalar mass should
be much larger than the physical pion mass; therefore, the early
change in the character of the transition appears as an
unrealistic feature of the logarithmic Polyakov potential. In the
case of the polynomic potential the onset of this first order
phase transition occurs for a pion mass larger than 700 MeV,
i.e.~in an energy region where the applicability of the effective
quark models is limited. This situation is qualitatively similar
to that observed in the local PNJL model.

In Fig.~\ref{fig3} we show the results for the mass dependence of the
critical transition temperatures within our nonlocal models. For comparison
we also quote typical curves obtained in the framework of the local PNJL
model (here we have considered the parameterization in
Ref.~\cite{Roessner:2006}). Upper and lower panels correspond to polynomic
and logarithmic Polyakov potentials, respectively, with $T_0=270$~MeV.
Before discussing in detail the results obtained for the nlPNJL models, let
us comment those corresponding to the PNJL model: from Fig.~\ref{fig3} we
observe that already at the physical value $m_\pi = m_\pi^{\rm phys}$ the
model predicts a noticeable splitting between the chiral restoration
temperature $T_{\rm ch}$ (dashed line) and the deconfinement temperature
$T_{\rm PL}$ (dotted line). In addition, it is seen that the growth of
$T_{\rm ch}$ with $m_\pi$ is stronger than that of $T_{\rm PL}$, which
implies that the splitting between both critical temperatures becomes larger
if $m_\pi$ is increased. This is not supported by existing lattice
results~\cite{Karsch:2000kv,Bornyakov:2009qh}, which indicate that both
transitions take place at approximately the same temperature, up to values
of $m_\pi$ even larger than those considered here. Comparing both panels it
is seen that the splitting is more pronounced for the PNJL model that
includes a logarithmic Polyakov potential.

We turn now to the curves obtained within nonlocal models. First of all,
from the figure it is seen that both parameterizations S1 and S2 lead to
qualitatively similar results. Contrary to the situation in the PNJL model,
in nlPNJL models both the chiral restoration and deconfinement transitions
occur at basically the same temperature for all considered values of
$m_\pi$. Moreover, comparing the results for the two alternative Polyakov
loop potentials we see that the main qualitative difference between them is
the already mentioned fact that in the case of the logarithmic potential
there is a critical pion mass of about 400 MeV where the character of the
transition changes from crossover to first order (dashed-dotted and solid
lines in the lower panel of Fig.~\ref{fig3}, respectively). By analyzing in
more detail the pion mass dependence of the critical temperatures, it is
seen that for $m_\pi$ above the physical mass the nlPNJL model results can
be accurately adjusted through a linear function
\begin{equation}
T_c (m_\pi) =  A \ m_\pi + B \ .
\end{equation}
This is in agreement with the findings of the lattice calculations of
Refs.~\cite{Karsch:2000kv,Bornyakov:2009qh}. Our results for the slope
parameter $A$ for both parameterizations and Polyakov loop potentials are in
the range of $0.06-0.07$. For comparison, most lattice calculations find $A
\lesssim
0.05$~\cite{Karsch:2007dt,Karsch:2000kv,Bornyakov:2005dt,Cheng:2006qk},
while according to some recent analyses~\cite{Ejiri:2009ac,Bornyakov:2009qh}
the value could be somewhat above this bound. Thus the slope parameter
predicted by the nonlocal PNJL models appears to be compatible with lattice
estimates. This can be contrasted with the results obtained within pure
chiral models, where one finds a strong increase of the chiral restoration
temperature with $m_\pi$~\cite{Berges:1997eu,Dumitru:2003cf,Braun:2005fj}.
For example, within the chiral quark model of Ref.~\cite{Braun:2005fj} one
gets a value $A=0.243$.

It is also worth to discuss the effect of considering a value of $T_0$ that
depends on the current quark masses, as suggested in
Ref.~\cite{Schaefer:2007pw}. The change in $T_0$ leads to an overall
decrease of the transition temperatures, which keep the rising linear
dependence on $m_\pi$ but with a slope parameter that gets reduced by about
$15-20$\%. The main noticeable difference is that in all cases the
transition becomes steeper, which leads to an earlier onset of the first
order transition. For example, for the parameter set S2 we find that the
transition becomes of first order already at $m_\pi \simeq 500$ MeV in the
case of the polynomic Polyakov potential, and about one half of this value
for the logarithmic one. These critical masses appear to be too small in
comparison with present lattice QCD estimations. However, in this respect it
is important to recall the importance of considering corrections that go
beyond the mean field approximation used here. In fact, although the role of
these corrections is expected to be less important as the quark mass
increases~\cite{Blaschke:2007np}, in the range of masses considered here
they can be significant enough to soften the transitions and lower the
critical temperatures~\cite{Horvatic:2010md}. In this sense, although a
fully nonperturbative scheme to account for meson fluctuations in nonlocal
models is still lacking, some important steps have been
taken~\cite{Blaschke:2007np,Hell:2008cc,Radzhabov:2010dd}. As it is pointed
out in Ref.~\cite{Benic:2012ec}, these fluctuations could also help to avoid
thermodynamical instabilities that could arise in nonlocal models.

\section{Summary and conclusions}

In this work we have analyzed the dependence of the deconfinement
and chiral restoration critical temperatures on the explicit
chiral symmetry breaking driven by the current quark mass. We work
in the framework of SU(2) nonlocal chiral quark models with
Polyakov loop (nlPNJL models), considering two different
functional forms of the Polyakov loop effective potential commonly
used in the literature, namely a polynomic function and a
logarithmic function. As a first step we have considered the mass
dependence of the pion mass and decay constant at vanishing
temperature, in comparison with that obtained in the local NJL
model and in lattice QCD. We have found that, while lattice
results for the ratio $m_\pi^2/m$ are in agreement with both local
and nonlocal models, those for $f_\pi$ show a significant increase
with $m_\pi$ that can be reproduced only by the predictions of
nonlocal models. Concerning the deconfinement and chiral
restoration critical temperatures, we have found that, contrary to
the case of the local PNJL model, in nlPNJL models both critical
temperatures turn out to be strongly entangled for the considered
range of pion masses. In addition, it is seen that the growth of
critical temperatures with the pion mass above the physical point
is basically linear, with a slope parameter which is close to
existing lattice QCD estimates. On the other hand, particularly in
the case of the logarithmic Polyakov loop potential,
the present mean field calculation leads to a too early onset of
the first order transition known to exist in the large quark mass limit.
We expect that the development of a fully
nonperturbative scheme to account for meson fluctuations
in nonlocal models might help to provide a solution to this
problem.

\begin{figure}[hbt]
\includegraphics[width=0.7 \textwidth]{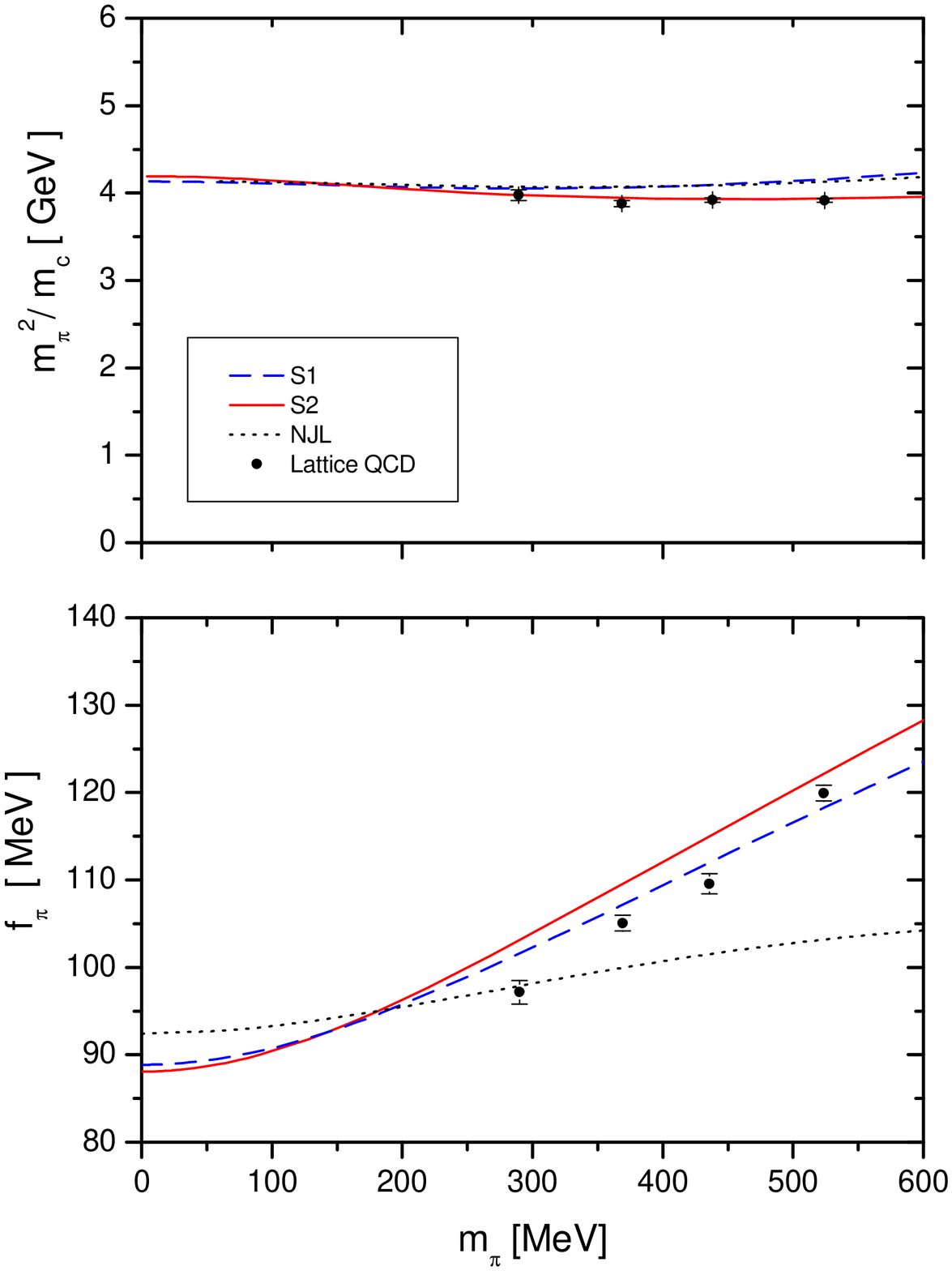}
\caption{Pion properties at $T=0$ as functions of the pion mass in local and
nonlocal chiral quark models. Upper an lower panels correspond to the ratio
$m_\pi^2/m_c$ and the pion decay constant $f_\pi$, respectively. Lattice
results are taken from Ref.~\cite{Noaki:2008iy} } \label{fig1}
\end{figure}

\begin{figure}[hbt]
\includegraphics[width=1. \textwidth]{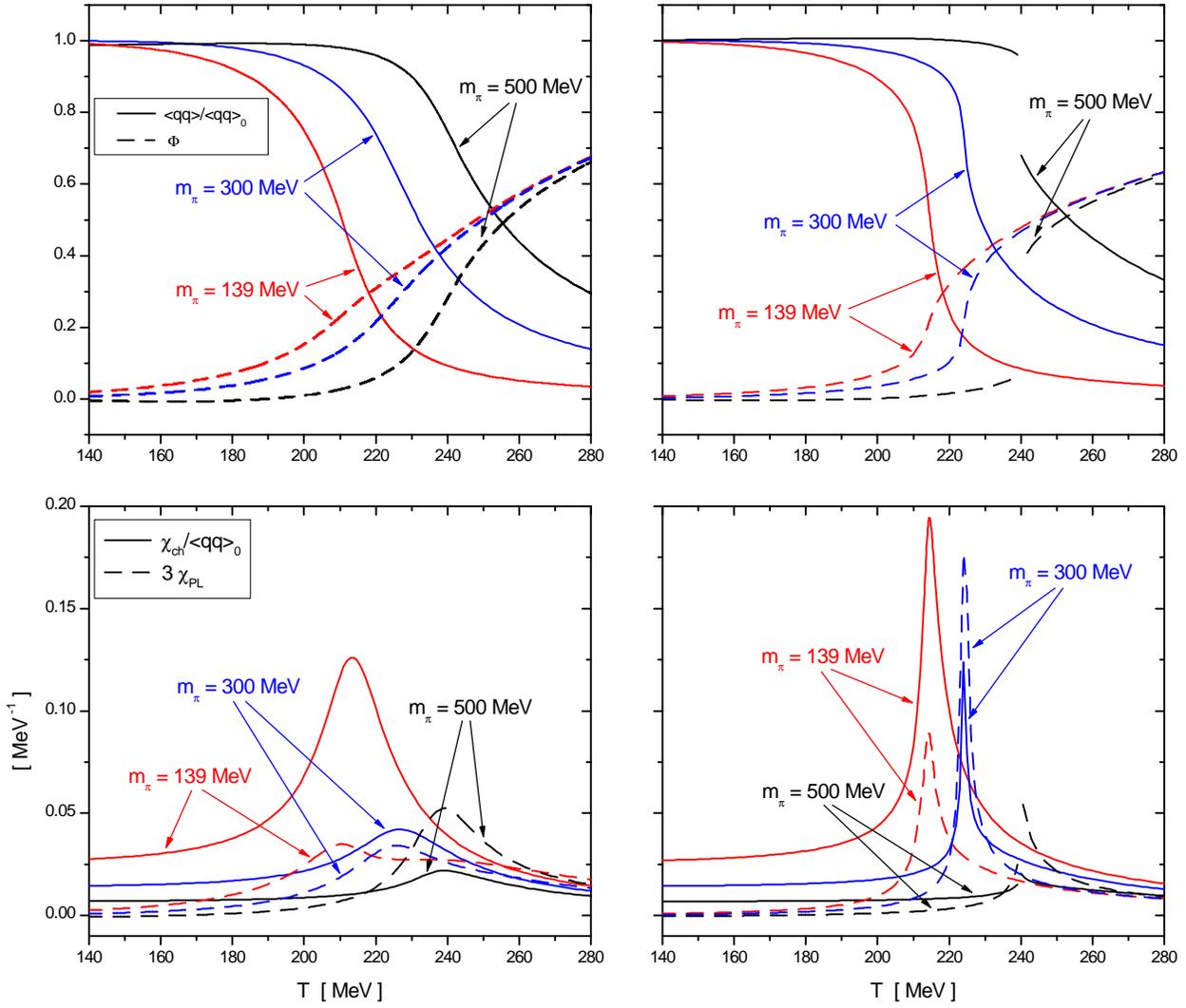}
\caption{Order parameters (upper panels) and the corresponding
susceptibilities (lower panels) as functions of the temeprature for some
representative values of the pion mass. Left (right) panels correspond to
the polynomic (logarithmic) Polyakov potential.} \label{fig2}
\end{figure}

\begin{figure}[hbt]
\includegraphics[width=0.7 \textwidth]{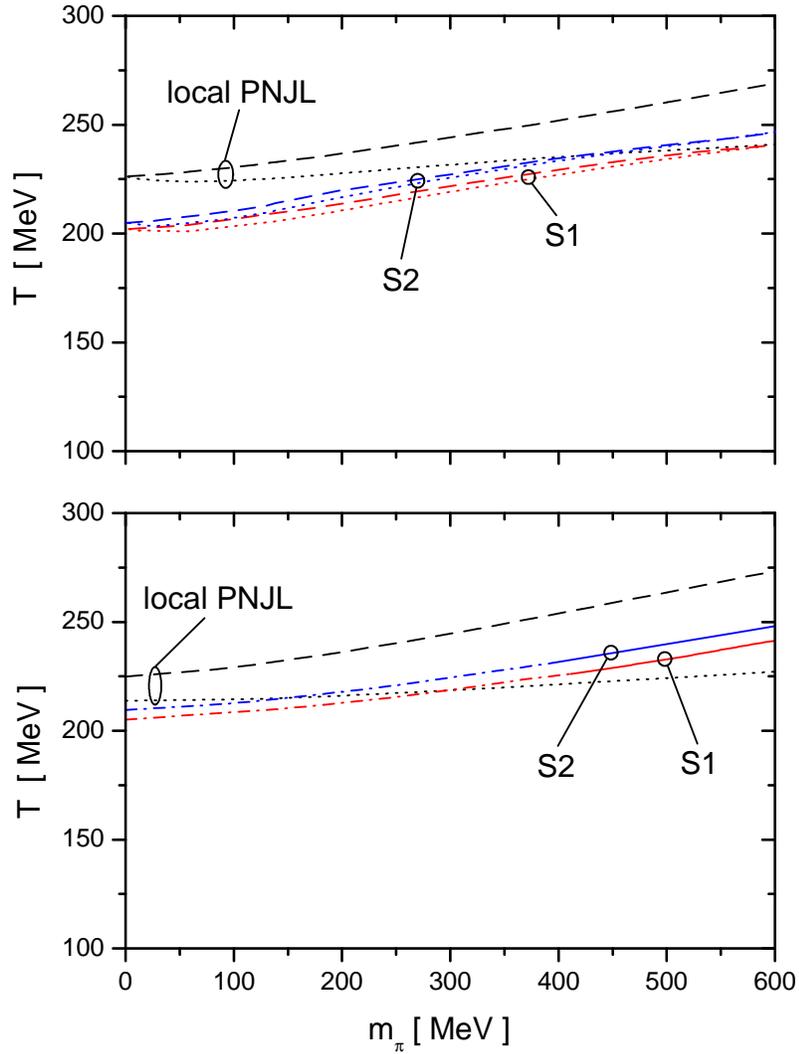}
\caption{Critical temperatures as functions of the pion mass for PNJL and
nlPNJL models, considering polynomic (upper panel) and logarithmic (lower
panel) Polyakov loop potentials. Dashed and dotted lines correspond to
chiral restoration and deconfinement transition temperatures, respectively.
For the nlPNJL models with a logarithmic potential (lower panel), both
transitions occur at the same temperature, and they can be of first order
(solid lines) or proceed as a smooth crossover (dashed-dotted lines).}
\label{fig3}
\end{figure}

\end{document}